\documentclass[%
reprint,
superscriptaddress,
amsmath,amssymb,
aps,
]{revtex4-2}
\usepackage{graphicx}
\usepackage{dcolumn}
\usepackage{bm}
\usepackage{xcolor}
\usepackage{siunitx}
\usepackage{array}
\usepackage{booktabs}
\usepackage{multirow}
\usepackage{dcolumn}

\begin{document}

\title{Time crystals and nonequilibrium dissipative phase transitions mediated by squeezed bath}
\author{Zhenghao Zhang}
\email{zzh312@tamu.edu}
\author{Qingtian Miao}
\email{qm8@tamu.edu}
  \affiliation{Department of Physics and Astronomy, Texas A\&M University, Texas 77843, USA}
\author{G. S. Agarwal}
\email{Girish.Agarwal@ag.tamu.edu}
 \affiliation{Institute for Quantum Science and Engineering, Department of Biological and Agricultural Engineering, Department of Physics and Astronomy, Texas A\&M University, College Station, Texas 77843, USA}

\date{\today}

\begin{abstract}
Nonequilibrium dissipative phase transition, arising from the competition of cooperative behavior and coherent field driving, discovered in the 1970s by Narducci et al. \cite{narducci1978transient} and Walls et al. \cite{walls1978non}, has been found to exhibit time-crystal behavior when the driving field exceeds the cooperative decay rate. This was seen through the study of the eigenvalues of the Liouvillian superoperator that describes the joint effect of drive and cooperativity. The cooperative decay depends on the nature of the reservoir correlations. If the reservoir correlations have phase-sensitive behavior, then the eigenvalues of the Liouvillian will be different. We investigate the time-crystal behavior of the nonequilibrium dissipative phase transitions under the influence of a squeezed vacuum reservoir. We analyze the steady-state phase diagram as a function of the control parameter and demonstrate that increasing the squeezing strength sharpens the dissipative phase transition. Spectral analysis of the Liouvillian reveals gap closings and the emergence of purely imaginary eigenvalues in the thermodynamic limit, indicating the time-crystal phase. We find that the real parts of subleading eigenvalues exhibit nonmonotonic behavior with increasing squeezing, reflecting the sensitivity of relaxation dynamics to the reservoir properties. Time-domain simulations confirm that the oscillation frequencies correspond to the imaginary parts of the Liouvillian eigenvalues. We also present results on quantum fluctuations in the time-crystal phase. Our results call attention to the study of time crystals in models of cooperativity based on engineered environments.
\end{abstract}

\maketitle

\section{Introduction}

The concept of time crystals—phases of matter characterized by the spontaneous breaking of time-translation symmetry—has evolved significantly since its theoretical inception \cite{PhysRevLett.109.160401, PhysRevLett.109.160402}. While no-go theorems established that such phases are forbidden in equilibrium systems with short-range interactions \cite{PhysRevLett.111.070402, PhysRevLett.114.251603}, the exploration of nonequilibrium systems has opened new avenues. In particular, periodically driven (Floquet) systems have provided fertile ground for discrete time crystals (DTCs) \cite{RevModPhys.95.031001, sacha2015modeling, PhysRevLett.117.090402, PhysRevLett.116.250401, PhysRevLett.118.030401, yao2020classical}. In such systems, the discrete time-translation symmetry imposed by the drive can be spontaneously broken, resulting in observables that oscillate with a period different from that of the drive. Experimental realizations of DTCs have been reported across a range of platforms, including trapped ions \cite{zhang2017observation, kyprianidis2021observation}, nuclear spins \cite{choi2017observation, rovny2018observation, o2020signatures, randall2021many, stasiuk2023observation, he2025experimental}, superfluid \cite{smits2018observation, autti2018observation}, and superconducting qubits \cite{frey2022realization, mi2022time}. 

More recently, the study of time crystals has extended to open quantum systems \cite{iemini2018boundary,hannukainen2018dissipation,kessler2019emergent,buvca2019non,piccitto2021symmetries,prazeres2021boundary,passarelli2022dissipative,cabot2023quantum,mattes2023entangled,krishna2023measurement,cabot2024continuous,yang2025emergent}, where engineered dissipation—rather than Floquet driving—plays a central role in enabling time-translation symmetry breaking. Within this context, continuous time crystals (CTCs) emerge as nonequilibrium steady states that exhibit persistent oscillations under time-independent Hamiltonians and dissipation. A paradigmatic model consists of a collection of driven two-level atoms interacting collectively with a bosonic mode under Markovian dissipation \cite{iemini2018boundary}. Above a critical drive strength, the system undergoes a dissipative phase transition, with the collective polarization entering a regime of sustained oscillations whose lifetime diverges in the thermodynamic limit. Crucially, stabilizing the time-crystal phase requires a discrete symmetry, such as $\mathbb{Z}_2$, explicitly broken by dissipation, together with a strong symmetry associated with total angular momentum conservation \cite{piccitto2021symmetries}. Experimental signatures of dissipative CTCs have been observed in platforms including nonlinear optical cavities \cite{kongkhambut2022observation}, erbium-doped solids \cite{chen2023realization}, and thermal Rydberg ensembles \cite{wu2024dissipative}.

In parallel, growing theoretical interest has focused on squeezed vacuum reservoirs—engineered environments formed by correlated photon pairs—as a means of controlling open quantum systems and preparing nonclassical states \cite{agarwal1990cooperative,agarwal1989nonequilibrium,dalla2013dissipative,qin2018exponentially,you2018waveguide,chen2021shortcuts,bai2021generating,gutierrez2023dissipative,groszkowski2022reservoir}. Owing to their phase-sensitive nature, such reservoirs can amplify or suppress specific quadrature fluctuations, a property that has enabled experimental insights into fundamental processes such as spontaneous emission \cite{murch2013reduction} and resonance fluorescence \cite{toyli2016resonance}. The theoretical framework for dissipative spin squeezing has its roots in early studies of atoms interacting with squeezed light, which revealed cooperative effects and the generation of nonclassical spin states \cite{agarwal1990cooperative,agarwal1989nonequilibrium}. More recent proposals have explored squeezed-reservoir engineering as a route to stabilize bosonic squeezed states and induce entanglement in atomic ensembles \cite{gutierrez2023dissipative}, showing that phase-sensitive dissipation can qualitatively reshape nonequilibrium dynamics and alter steady-state behavior \cite{groszkowski2022reservoir}.

In this work, we extend the driven-dissipative collective spin model of Ref. \cite{iemini2018boundary} by coupling the atomic ensemble to a squeezed vacuum reservoir, drawing inspiration from early studies of cooperative emission irradiated by squeezed light \cite{agarwal1990cooperative}. While a recent experiment on the superconducting Duffing oscillator has demonstrated a first-order dissipative phase transition and identified long-lived metastable states through Liouvillian spectral analysis \cite{chen2023quantum}, our focus is on a second-order transition associated with the emergence of time-crystalline order. We begin by revisiting the standard dissipative phase transition in the case of a vacuum reservoir in Sec. \ref{DPTQ}, and then present the modified phase diagram under squeezed radiation in Sec. \ref{PDS}. We show that the presence of squeezing sharpens the phase transition. Spectral analysis of the Liouvillian in Sec. \ref{SpLiou} reveals that squeezing alters the closing of the spectral gap. In particular, the real parts of subleading eigenvalues exhibit nonmonotonic dependence on the squeezing parameter, reflecting the sensitivity of relaxation dynamics to reservoir correlations. Time-domain simulations in Sec. \ref{TiEvo} further confirm that the dominant oscillation frequencies correspond precisely to imaginary components of the Liouvillian spectrum. Finally, we explore the impact of squeezing on steady-state quantum fluctuations in Sec. \ref{Fluc}, using the Wigner function and population distributions to illustrate signatures of the underlying dissipative phase transition.

\section{Dissipative Phase Transition of Atoms Coupled to a Vacuum Mode in a Low-Q Cavity}
\label{DPTQ}

Dissipative phase transitions were first predicted in the 1970s for systems of $N$ coherently driven two-level atoms \cite{narducci1978transient,drummond1978volterra,walls1978non,puri1979exact,carmichael1980analytical} confined within a subwavelength volume, such that all atoms interact via a common, spatially uniform scattered field. A second-order phase transition emerges in the regime of collective damping, where $\langle J^2 \rangle$ is conserved, with $J$ the total pseudo-angular momentum. Similar cooperative decay can also arise in extended systems coupled to a single-mode field, under conditions analogous to single-mode superradiance. Such dynamics can be realized in a low-Q (bad) cavity, where photons leak out through partially transmitting mirrors at a rate $\kappa$. In the bad cavity limit, where $\kappa$ far exceeds both collective and individual atomic decay rates, the cavity field may be adiabatically eliminated, yielding effective atomic equations of motion. 

Under the Born-Markov approximation, and in the rotating frame of the resonant drive, the master equation for the reduced atomic density operator $\rho$ takes the form
\begin{equation}
\dot\rho=-i\Omega[S^++S^-,\rho]+2\Gamma\mathcal L_{S^-}\rho,
\label{lowq}
\end{equation}
where $\Omega$ is the Rabi frequency and $S^\pm = \sum_i S_i^\pm$ the collective raising and lowering operators acting as ladder operators for an effective spin-$N$/2 system. The polarization operator is given by $S^z=\sum_i S_i^z$. The Lindblad superoperator is defined by $\mathcal{L}_\xi \rho = \xi \rho \xi^\dagger - \frac{1}{2} \{ \xi^\dagger \xi, \rho \}$. The single-atom decay rate is $\Gamma = g^2/\kappa$, with $g$ the atom-field coupling strength. Since $\langle J^2 \rangle$ is conserved, the dynamics remain confined to the symmetric subspace with maximal total angular momentum. Note that the same effective master equation also describes an array of $N$ atoms coupled to a one-dimensional waveguide when the interatomic spacing is commensurate with the drive wavelength—that is, when the phase shift between adjacent atoms equals an integer multiple of $2\pi$.

Ref. \cite{narducci1978transient} shows that the normalized atomic polarization $\langle S^z\rangle/N$ varies smoothly and monotonically with the scaled drive strength $\Omega/\Omega_0$, where $\Omega_0=\Gamma N/2$, for any finite $N$. For $\Omega/\Omega_0<1$, this quantity evolves continuously from $-1/2$ (all atoms in the ground state) to $0$ (complete saturation) as $\Omega$ increases. When $\Omega/\Omega_0>1$, the collective spin states become equally populated. In the thermodynamic limit $N\rightarrow \infty$ with $\Omega/\Omega_0$ remaining finite, this sharpens into a second-order phase transition at the critical point $\Omega / \Omega_0 = 1$. Using a Fokker-Planck method, Ref. \cite{drummond1978volterra,walls1978non} further predict that above the critical point ($\Omega / \Omega_0 > 1$) and in the thermodynamic limit, the system no longer relaxes to a time-independent steady state. Instead, it evolves into a family of periodic, oscillatory solutions.

\begin{figure}[bp]
    \includegraphics[width=0.48\textwidth]{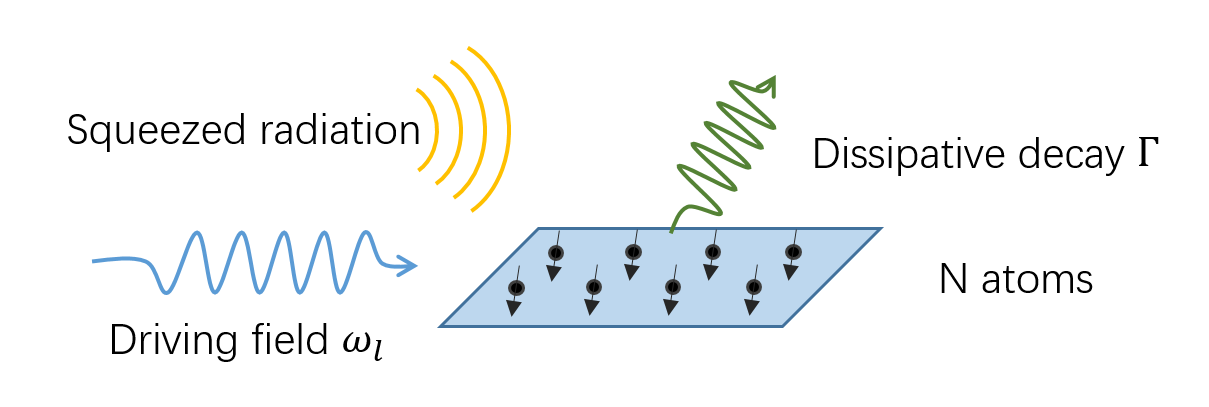}
    \caption{Schematic of the system: $N$ identical two-level atoms are resonantly driven by a coherent field at frequency $\omega_l$ and coupled to a broadband squeezed vacuum reservoir. The coupling to the electromagnetic environment is characterized by the spontaneous emission rate $\Gamma$. The squeezed radiation modifies the quantum fluctuations of the electromagnetic field by an effective photon occupation number $\bar n$ and a two-photon correlation amplitude $m$.}
    \label{model}
\end{figure}

\begin{figure*}[tp]
    \includegraphics[width=0.95\textwidth]{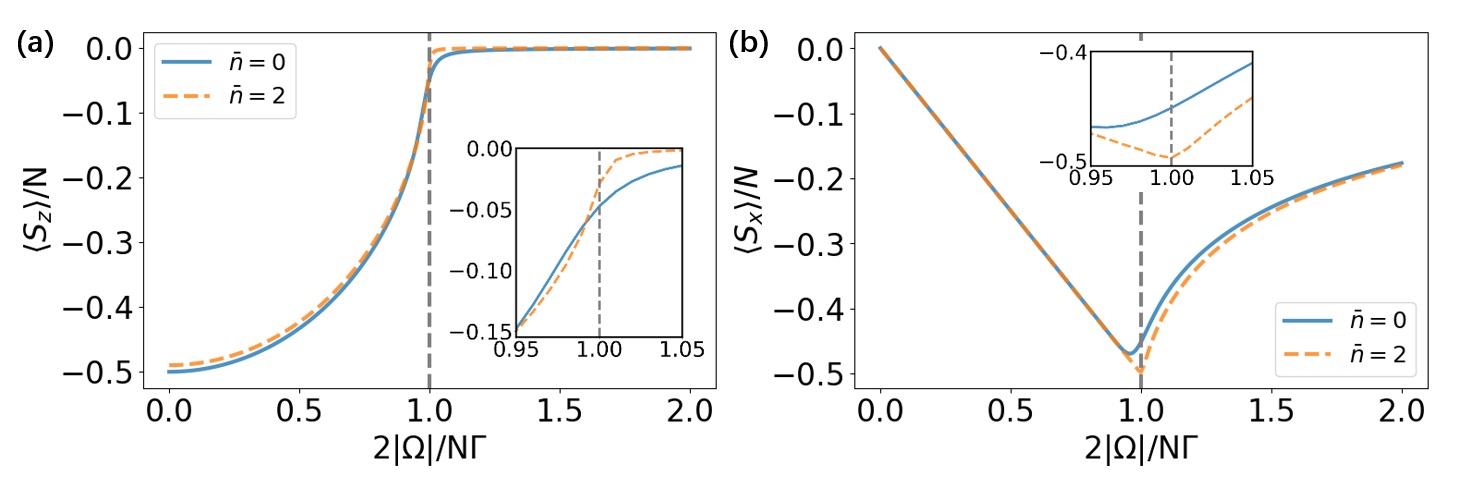} 
    \caption{Steady-state expectation values as a function of the scaled drive strength $2|\Omega|/N \Gamma$ for two values of the effective squeezing photon number, $\bar{n}=0$ and $\bar{n}=2$. (a) Longitudinal polarization $\left < S_z \right >$, (b) transverse polarization $\left < S_x \right >$. Insets show magnified views near the critical point. The atom number is fixed at $N=100$.}
    \label{phase}
\end{figure*}

Dissipative phase transitions have attracted growing interest due to their connection with time-crystalline behavior \cite{iemini2018boundary,hannukainen2018dissipation,gegg2018superradiant,cabot2023quantum}. Although the right-hand side of Eq. (\ref{lowq}) is time-independent, in the thermodynamic limit and at asymptotically long times, $\langle S^z\rangle/N$ exhibits persistent oscillations with a period determined solely by the system's coupling parameters. This marks the spontaneous breaking of continuous time-translation symmetry, distinguishing a time-crystal phase from a stationary phase, separated by the critical point. The dissipative phase transition is characterized by the closing of the Liouvillian spectral gap, as discussed in Ref. \cite{iemini2018boundary}. For $\Omega / \Omega_0 < 1$, the Liouvillian spectrum is gapped, and the slowest decaying modes are purely real, ensuring convergence to a unique steady state. For $\Omega / \Omega_0 > 1$, the spectrum becomes gapless, and the dominant eigenvalues acquire nonzero imaginary parts, resulting in persistent, time-periodic dynamics in the thermodynamic limit.

\section{Dissipative phase transition in presence of squeezed radiation}
\label{PDS}

We consider a system of $N$ identical two-level atoms with transition frequency $\omega_0$, driven resonantly by a coherent field at frequency $\omega_l=\omega_0$, and coupled to a broadband squeezed vacuum reservoir, as in Fig. \ref{model}. To derive the master equation, we employ both the electric-dipole and rotating-wave approximations and temporarily drop the coherent drive. In this framework, all atoms interact identically with the local electromagnetic field, described by the interaction Hamiltonian (we set $\hbar=1$):
\begin{equation}
H_{SR}=\sqrt{2\Gamma}[\mathcal E_S(t) S^++\mathcal E_S^\dagger(t) S^-],
\end{equation}
where $\Gamma$ is the spontaneous emission rate into the unsqueezed vacuum, and $\mathcal E_S(t)$, $\mathcal E_S^\dagger(t)$ are the positive- and negative-frequency components of the squeezed field.

We consider squeezing generated by a broadband parametric amplifier that emits correlated photon pairs $a$ and $b$, satisfying the phase-matching condition $\omega_a+\omega_b=2\omega_c$, where $\omega_c$ is the central frequency of the amplifier. These correlations produce a squeezed vacuum in which field modes at $\omega_a$ and $\omega_b$ are entangled. We set $\omega_c=\omega_0$, resonant with the atomic transition, and assume the amplifier bandwidth is much broader than the atomic decay rate. This ensures that the squeezed vacuum encompasses all electromagnetic modes relevant to the atomic dynamics. Under this condition, the atoms interact exclusively with squeezed modes, and the field appears as $\delta$-correlated white noise \cite{gardiner1985input,carmichael1987resonance}:
\begin{equation}
\begin{aligned}
&\langle \mathcal E_S^\dagger(t)\mathcal E_S(t')\rangle=\bar{n}\delta(t-t'),\\
&\langle \mathcal E_S(t)\mathcal E_S^\dagger(t')\rangle=(\bar{n}+1)\delta(t-t'),\\
&\langle \mathcal E_S(t)\mathcal E_S(t')\rangle=e^{-2i\omega_c t}m\delta(t-t')\\
&\langle \mathcal E_S^\dagger(t)\mathcal E^\dagger_S(t')\rangle=e^{2i\omega_c t}m^*\delta(t-t'),
\end{aligned}
\end{equation}
and $\langle\mathcal E_S(t)\rangle=0$, where $\bar{n}$ and $m$ quantify the photon number and two-mode correlations of the squeezed state. These parameters satisfy the inequality $|m|^2\le \bar n(\bar n+1)$, with equality for minimum uncertainty squeezing.

We adopt the Born and Markov approximations and derive the master equation in the interaction picture with respect to the free atomic Hamiltonian $\omega_0 S^z$, including the effect of the coherent drive. Following the standard procedure and setting perfect squeezing $|m|^2 = \bar n(\bar n+1)$, where $m=|m|e^{i\phi}$ and we take $\phi=0$ throughout, the master equation for the atomic density matrix $\rho$ takes the form \cite{agarwal1990cooperative}:
\begin{equation}
\begin{aligned}
    \dot \rho &=  i[|\Omega| \exp(i \psi) S^+ + |\Omega| \exp(-i \psi) S^-, \rho] \\
    &- \Gamma(1+\bar{n})(S^+ S^- \rho - 2 S^- \rho S^+ + \rho S^+S^-) \\ 
    &- \Gamma \bar{n} (S^- S^+ \rho - 2 S^+ \rho S^- + \rho S^-S^+) \\ 
    &- \Gamma |m| (S^+ S^+ \rho - 2 S^+ \rho S^+ + \rho S^+S^+) \\ 
    &- \Gamma |m| (S^- S^- \rho - 2 S^- \rho S^- + \rho S^-S^-).
\end{aligned}
\label{rhodot}
\end{equation}
Here, the complex Rabi frequency is defined as $\Omega=|\Omega|e^{i\psi}$, with $\psi=\pi/2$ fixed. 

\begin{figure*}[htbp]
    \includegraphics[width=0.95\textwidth]{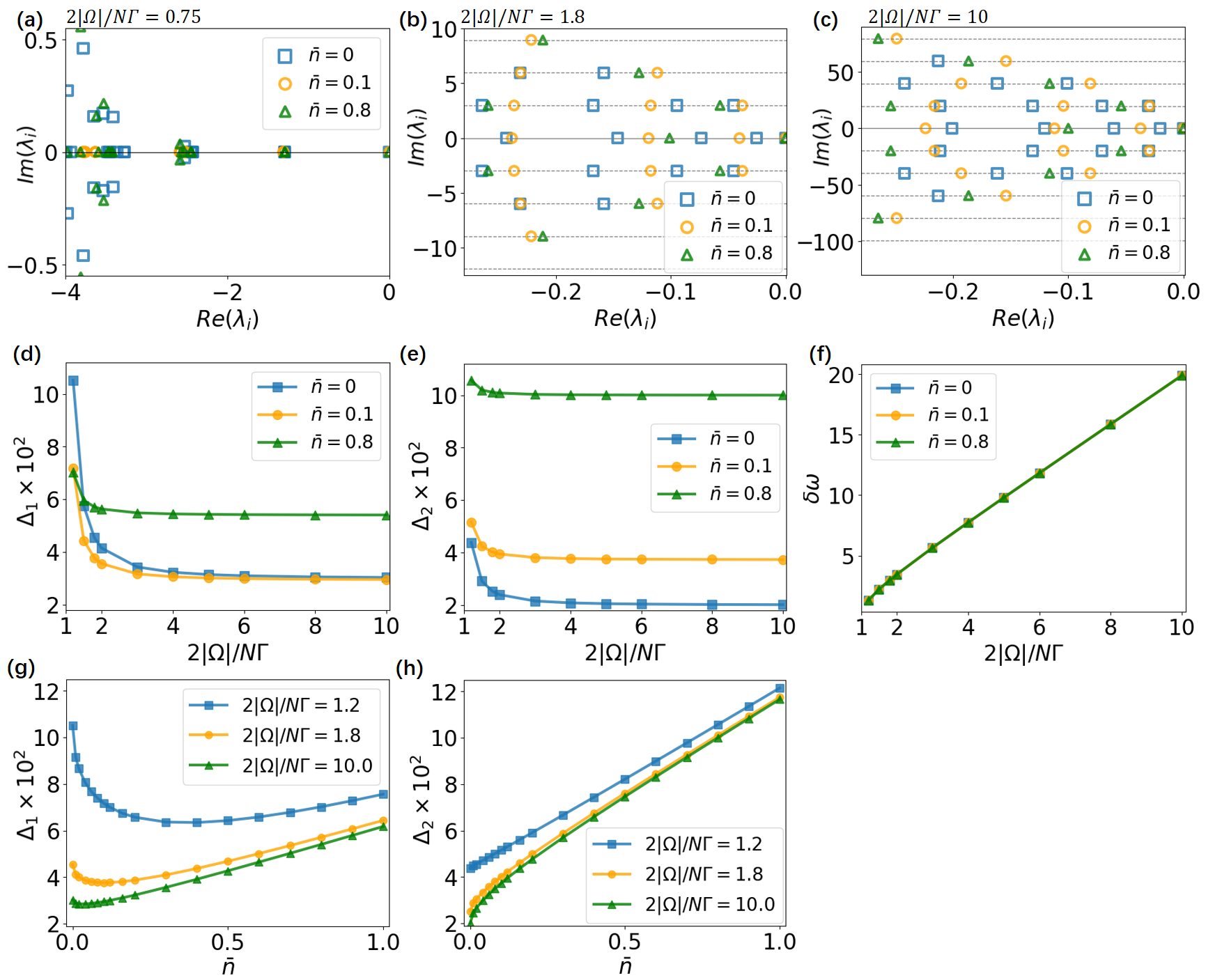} 
    \caption{Liouvillian spectra for $2|\Omega|/N\Gamma$ equals to (a) $0.75$, (b) $1.8$. (c) $10$. Panel (a) corresponds to the stationary phase, while (b) and (c) lie in the time-crystalline phase. Panels (d) and (e) show dissipative gaps $\Delta_{1}$ and $\Delta_2$, defined as the absolute values of the real parts of the first nonzero eigenvalues with nonzero and zero imaginary components, respectively, plotted versus $2|\Omega|/N\Gamma$ for $\bar n = 0,0.1,0.8$. Panel (f) shows $\delta\omega$, the absolute value of the first nonzero imaginary part, also as a function of $2|\Omega|/N\Gamma$ for the same values of $\bar n$. Panels (g) and (h) show $\Delta_{1}$ and $\Delta_2$, respectively, as functions of $\bar n$ for $2|\Omega|/N\Gamma=1,2,1.8,10$. The eigenvalues are plotted in units of $N\Gamma/2$. The other parameter used is $N=100$.}
    \label{Liou}
\end{figure*}

To investigate the impact of squeezing on the dissipative phase transition discussed in Sec. \ref{DPTQ}, we compute steady-state observables for a finite ensemble of $N=100$ atoms. In the thermodynamic limit, the transition is marked by a non-analytic change in observables; this sharp behavior becomes a gradual transition for finite systems. We focus on the longitudinal and transverse polarizations, $\langle S_z \rangle /N$ and $\langle S_x \rangle /N$, as functions of the scaled drive strength $2|\Omega|/N\Gamma$, which delineates a strong dissipative regime ($2|\Omega|/N\Gamma<1$) and a weak dissipative regime ($2|\Omega|/N\Gamma>1$). 

Figure \ref{phase}(a) shows that the longitudinal polarization $\langle S_z \rangle /N$ increases toward zero in both the unsqueezed ($\bar n =0$) and squeezed ($\bar n =2$) cases as the system crosses the critical point. However, for squeezed input, the rise in $\langle S_z \rangle /N$ near the critical point is steeper than in the unsqueezed case, indicating a sharper transition. Although $\langle S_z \rangle /N$ does not reach zero in the weak dissipative regime due to finite-size effects, it remains small (on the order of $10^{-3}$) and approaches zero from below with increasing $\bar n$, as shown in Fig. \ref{evol}(a).

The transverse polarization, shown in Fig. \ref{phase}(b), decreases toward $-1/2$ in the strong dissipative phase and begins to increase near the critical point. Notably, for $\bar{n}=2$, the minimum occurs close to $2|\Omega|/N\Gamma=1$, while for $\bar n =0$, the upturn begins slightly earlier. The component $\langle S_y \rangle /N$ remains zero across the full parameter range (not shown). These results demonstrate that squeezed radiation sharpens the transition and enhances the contrast between the two phases.

\section{Spectrum of Liouvillian and time crystalline phase}
\label{SpLiou}

Introducing the Liouvillian superoperator $\mathcal{L}$, the master equation in Eq. (\ref{rhodot}) can be written compactly as $\dot \rho = \mathcal{L} \rho$. Its formal solution takes a form analogous to unitary evolution under a Hamiltonian:
\begin{equation}
    \rho(t)  = e^{\mathcal{L}t}  \rho(0)  = \sum_i c_i e^{\lambda_i t} \rho_i,
    \label{evol}
\end{equation}
where $\lambda_i$ denotes the $i$th eigenvalue of the Liouvillian superoperator, $\rho_i$ are the corresponding right eigenmodes, and the coefficients $c_i$ are determined by the initial state. Solving the Liouvillian spectrum provides insight into the dynamical behavior of the system. In a pump-decay system, the density matrix $\rho(t)$ remains finite in the long-time limit, implying $\mathrm{Re}(\lambda_i) \leq 0$ for all $i$. By the Brouwer fixed-point theorem, at least one eigenvalue is expected to be zero, corresponding to the non-oscillating steady state \cite{watrous2018theory}. Expressing the time evolution in exponential form reveals that the negative real parts of the eigenvalues govern the decay rates of the associated modes, whereas the imaginary parts determine their oscillation frequencies.

The closing of the Liouvillian gap provides a clear signature of a dissipative phase transition. Figures \ref{Liou}(a–c) show the Liouvillian spectra for different values of drive strength and squeezing. In the strong dissipative regime [Fig. \ref{Liou}(a)], the spectrum is gapped, with the leading nonzero eigenvalues purely real and relatively large in magnitude, independent of squeezing. The dynamics in this regime are dominated by fast-decaying, non-oscillatory modes, while any subleading oscillatory modes decay more rapidly. In the weak dissipative regime [Fig. \ref{Liou}(b,c)], the real parts of the leading eigenvalues approach zero, indicating the closure of the Liouvillian gap. Simultaneously, the eigenvalues acquire imaginary nonzero components, marking the onset of long-lived oscillatory dynamics associated with time-crystalline behavior. While strictly non-decaying oscillations require the thermodynamic limit, the appearance of low-lying complex eigenvalues in finite systems already reflects time-crystal-like dynamics. Notably, for $\bar{n}=0$, the second eigenvalue remains purely real, whereas the introduction of squeezing induces a nonzero imaginary part, signaling the emergence of oscillatory modes.

\begin{figure}[htbp]
    \includegraphics[width=0.48\textwidth]{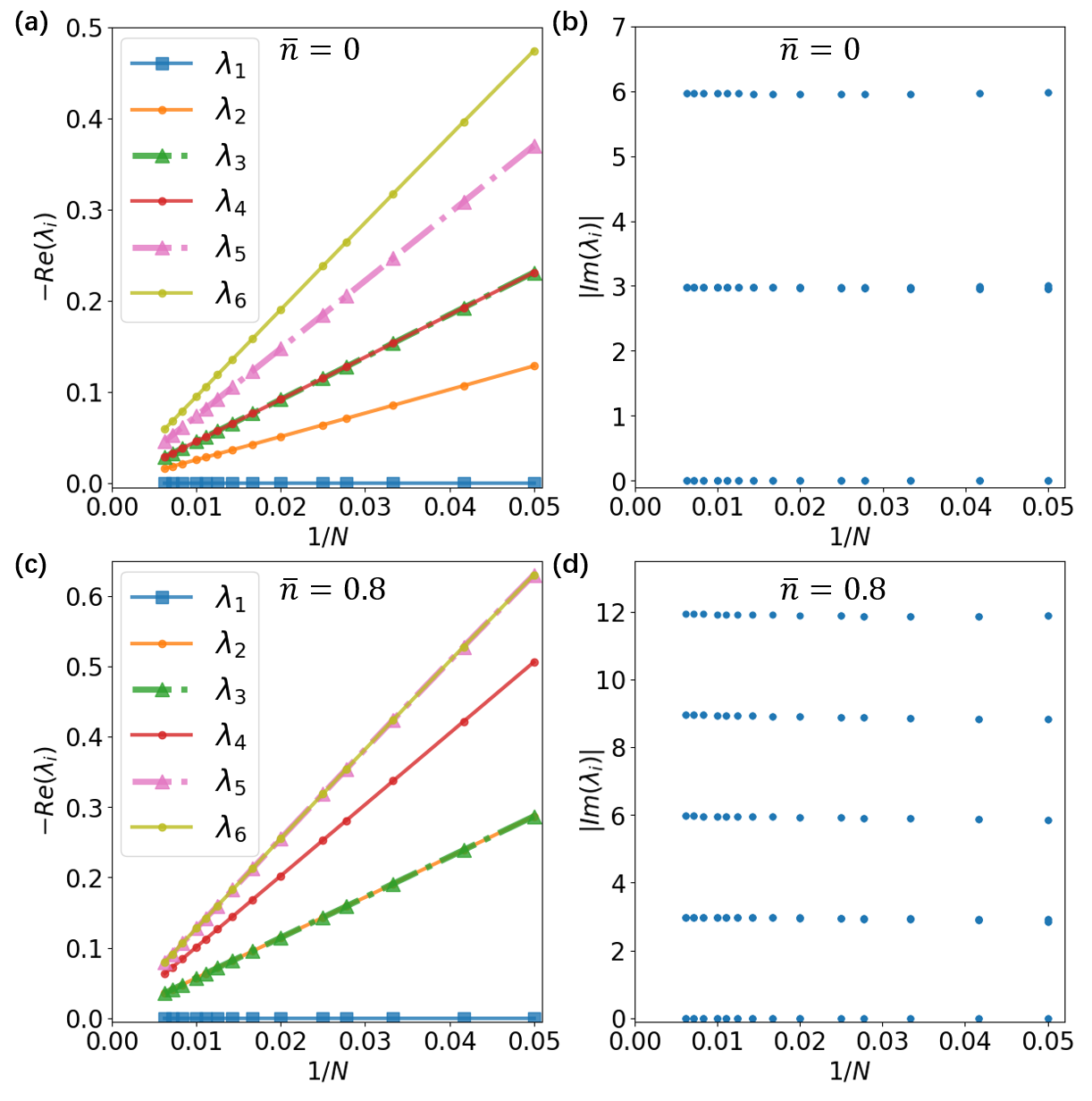}
    \caption{Finite size scaling for the real and imaginary parts of the Liouvillian eigenvalues in the time-crystal phase. (a-d) Eigenvalues plotted as a function of $1/N$ for (a,b) $\bar{n}=0$ and (c,d) $\bar{n}=0.8$. The eigenvalues are ordered by increasing absolute value of their real parts, i.e., $|{\rm Re}(\lambda_i)|<|{\rm Re}(\lambda_{i+1})|$. The absolute values of the first 6 real and 12 imaginary parts are shown, with eigenvalues expressed in units of $N\Gamma/2$. The other parameter used is $2|\Omega|/N\Gamma=1.8$. In the real part plot (a,c), overlapping lines indicate eigenvalues with imaginary components (forming conjugate pairs), while single lines correspond to purely real eigenvalues.}
    \label{eigen}
\end{figure}

To further explore the time-crystalline regime, we maintain a constant decay parameter $N\Gamma/2$ while gradually increasing the pumping strength $|\Omega|$. Figures \ref{Liou}(d–f) present the corresponding Liouvillian spectra. We define two dissipative gaps: $\Delta_{1,2}$, as the absolute values of the real parts of the first nonzero eigenvalues with nonzero and zero imaginary components, respectively. The spectral bandwidth $\delta\omega$ is defined as the absolute value of the first nonzero imaginary part. As shown in Fig. \ref{Liou}(d,e), both $\Delta_1$ and $\Delta_2$ decrease with increasing $|\Omega|$, eventually converging to finite values due to finite-size effects. Notably, Fig. \ref{Liou}(d) shows a crossing point: for moderate driving just beyond the critical point, stronger squeezing (e.g. $\bar{n}=0.8$) yields a smaller gap than the unsqueezed case, indicating longer-lived oscillatory modes; however, as the drive strength increases further, the gap for $\bar n = 0.8$ eventually exceeds that of the unsqueezed case. This crossover highlights the nonmonotonic influence of squeezing on the spectral gap. 

Fig. \ref{Liou}(f) shows that the frequency gap $\delta\omega$ grows linearly with the pump strength and remains independent of the squeezing parameter $\bar n$ over the range shown. The effect of squeezing on the dissipative gaps is further illustrated in Figs. \ref{Liou}(g,h), where $\Delta_{1}$ and $\Delta_2$ are plotted as functions of $\bar n$ at fixed driven strength $|\Omega|$. The curve of $\Delta_1$ versus $\bar{n}$ exhibits a nonmonotonic, U-shaped dependence, reaching a minimum value $\Delta_{1,min} \approx 0.0377$ at $\bar{n}=0.1$ for $2|\Omega|/N\Gamma=1.8$. As the drive increases, the location of this minimum shifts toward smaller $\bar{n}$. In contrast, $\Delta_2$ increases monotonously with $\bar{n}$. Comparing Fig. \ref{Liou} (g) and (h), $\Delta_2$ is initially smaller than $\Delta_1$; but surpasses it as $\bar{n}$ increases. This crossover further emphasizes the nontrivial role of squeezing in shaping the spectral structure of the Liouvillian.

To investigate the Liouvillian spectrum at larger system sizes, we analyze the real and imaginary components of the eigenvalues separately as functions of $1/N$. Figure \ref{eigen} shows the absolute values of the $6$ smallest nonzero real parts and the first $12$ nonzero imaginary parts of the Liouvillian eigenvalues, ordered such that $|{\rm Re}(\lambda_{i})| < |{\rm Re}(\lambda_{i+1})|$, with $\lambda_{1}=0$ corresponding to the non-oscillating steady state. We take all the eigenvalues into account, including ones with non-zero and zero imaginary parts. All eigenvalues are included, regardless of whether their imaginary parts vanish. In the real-part plots, eigenvalues with nonzero imaginary components appear as overlapping curves, while nondegenerate curves correspond to purely real eigenvalues. This structure reflects the spectral symmetry illustrated in Fig. \ref{Liou}. Although we examine scaling in $1/N$, we retain $N\Gamma/2$ as the unit of frequency throughout. Accordingly, $\Gamma$ is scaled inversely with $N$ to keep the effective collective decay rate fixed, following the convention of \cite{iemini2018boundary}. 

Figures \ref{eigen}(a,c) show the real parts of the Liouvillian eigenvalues as functions of $1/N$, with and without squeezing. The results exhibit an approximately linear scaling in $1/N$. In both cases, eigenvalues with nonzero imaginary parts have real components that tend toward zero in the thermodynamic limit ($N\rightarrow\infty$), indicating the emergence of non-decaying oscillatory modes. Their presence confirms time-crystalline behavior in both squeezed and unsqueezed regimes. Figures \ref{eigen}(b,d) show the corresponding imaginary parts of the eigenvalues, which form an approximately equally spaced band structure with characteristic spacing $\delta \omega$. As the squeezing parameter $\bar{n}$ increases, a greater number of long-lived oscillatory modes appear among the low-lying Liouvillian eigenstates, highlighting the role of squeezing in enhancing the spectral richness of the time-crystalline phase.

\section{Spectrum of Liouvillian and time evolution}
\label{TiEvo}

\begin{figure}[ht]
    \includegraphics[width=0.48\textwidth]{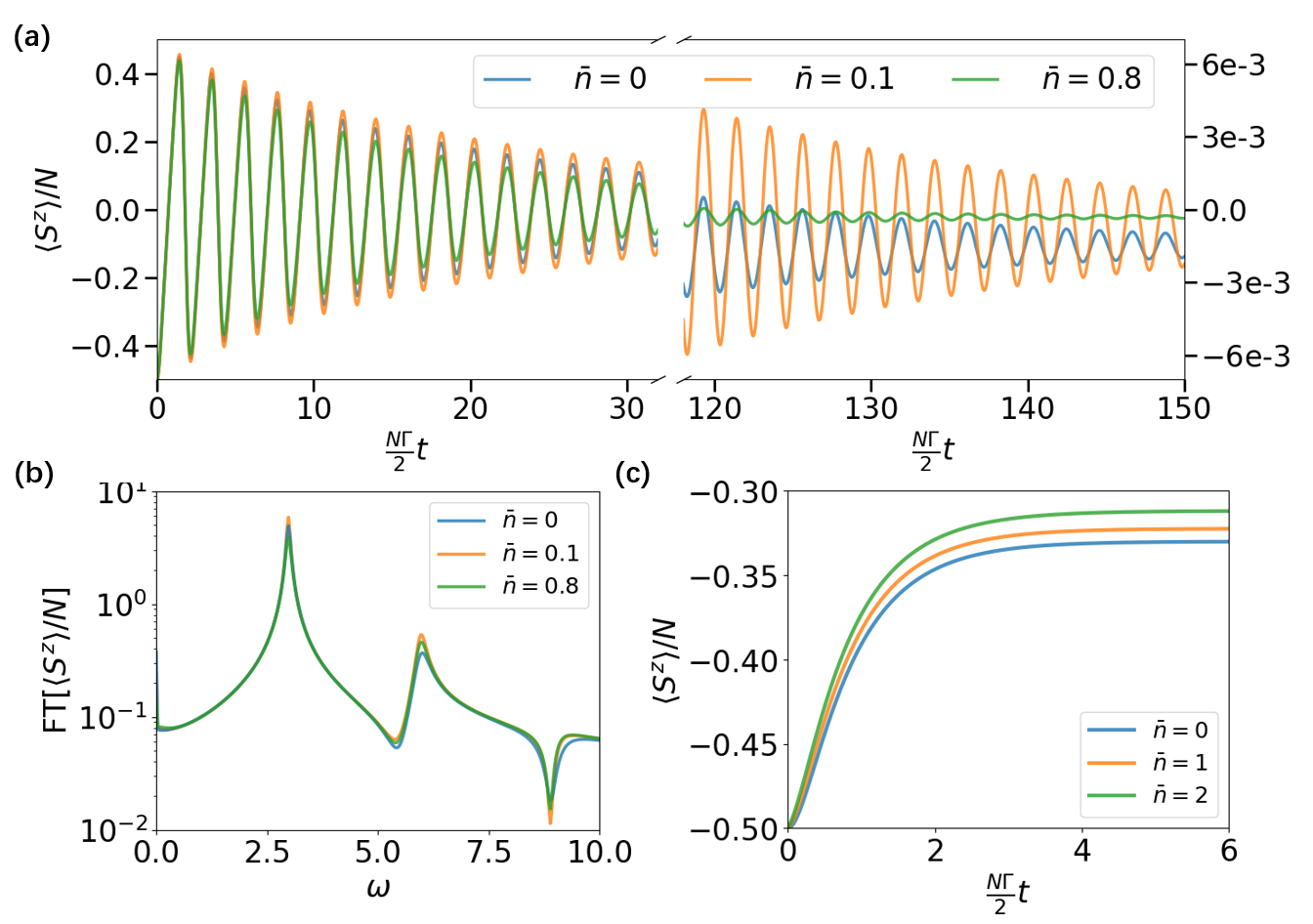}
    \caption{Time evolution of the normalized spin polarization $\langle S_z \rangle/N$. (a) Dynamics in the time-crystalline phase at $2|\Omega|/N\Gamma=1.8$. The left panel shows the initial evolution over $(N\Gamma/2)t\in[0,30]$; the right panel displays long-time behavior in the interval $(N\Gamma/2)t\in[120,150]$. (b) Fourier spectrum corresponding to the time trace in (a). (c) Evolution in the stationary phase at $2|\Omega|/N\Gamma=0.75$, showing fast decay to a non-oscillating steady state. The atom number is fixed at $N=100$. }
    \label{osci}
\end{figure}

To investigate the system’s dynamical behavior, we simulate the time evolution of the longitudinal polarization $\langle S^z \rangle / N$ for varying squeezing strengths, with a fixed atom number $N = 100$. The system is initialized in a pure state with all spins down. Figure \ref{osci}(a) shows the dynamics in the time-crystalline phase. Due to the finite system size, the oscillations exhibit gradual decay over time. The results are divided into early- and late-time regimes, where the oscillations are prominent and nearly vanished, respectively. The relaxation timescale is estimated as $\tau = 1/\Delta_1$. At higher squeezing levels (e.g., $\bar{n} = 0.8$), the oscillations decay rapidly, whereas at moderate squeezing (e.g., $\bar{n} = 0.1$), they persist significantly longer. The unsqueezed case ($\bar{n} = 0$) exhibits intermediate decay behavior, lying between these two scenarios. This non-monotonic trend is consistent with the U-shaped dependence of $\Delta_1$ on $\bar{n}$ shown in Fig. \ref{Liou}(g). At late times, the steady-state values of $\langle S^z \rangle / N$ settle around $10^{-3}$, decreasing with increasing squeezing strength. 

Figure \ref{osci}(b) presents the Fourier spectrum of the dynamics in panel (a). The first prominent peak appears at $2.98$, matching the spectral spacing observed in Fig. \ref{eigen}(b,d), thereby confirming that the oscillation frequencies are governed by the imaginary parts of the Liouvillian eigenvalues. For comparison, Fig. \ref{osci}(c) shows the system's behavior in the stationary phase, where the dynamics rapidly relax to a steady state without oscillations, as expected.

\section{Fluctuation Characteristics of the dissipative phase transition}
\label{Fluc}
\begin{figure}[htbp]
    \includegraphics[width=0.48\textwidth]{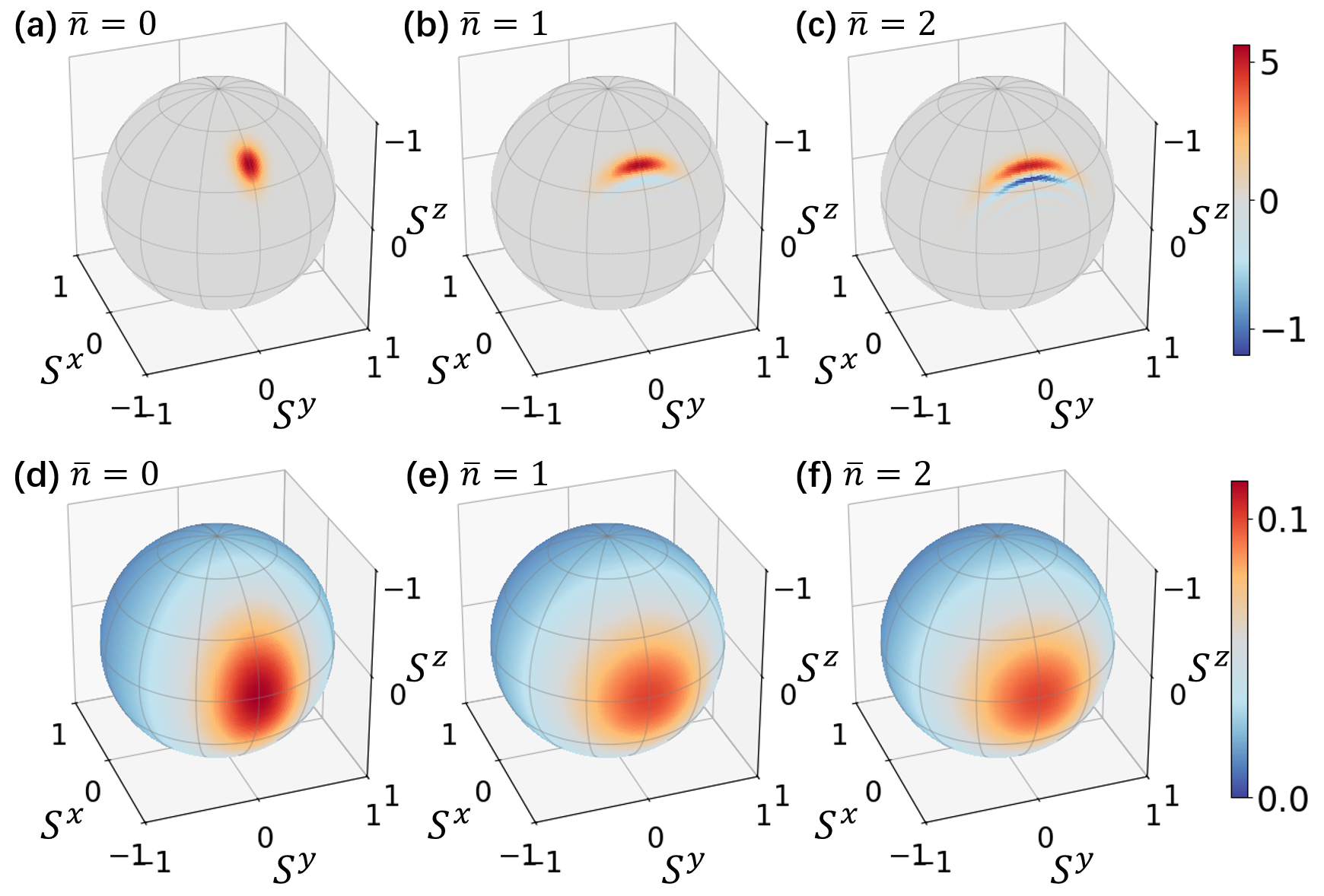}
    \caption{Bloch sphere heatmaps of the Wigner distribution for steady states at $N=100$ and varying $\bar{n}$. Panels (a–c) correspond to the stationary phase at $2|\Omega|/N \Gamma=0.75$, while (d–f) correspond to the time-crystalline phase at $2|\Omega|/N \Gamma=1.8$.}
    \label{Wigner}
\end{figure}
Quantum fluctuations remain a fundamental topic of interest in quantum optics, particularly in driven-dissipative systems. To characterize their behavior, we analyze the steady-state Wigner distributions under varying squeezing parameters, as shown in Fig. \ref{Wigner}, comparing both the stationary and time-crystalline phases. In the stationary phase, the Wigner distribution remains sharply localized on the surface of the Bloch sphere. In contrast, the time-crystalline phase exhibits a significantly broader distribution, indicative of enhanced quantum fluctuations.

\begin{figure}[hbtp]
    \includegraphics[width=0.48\textwidth]{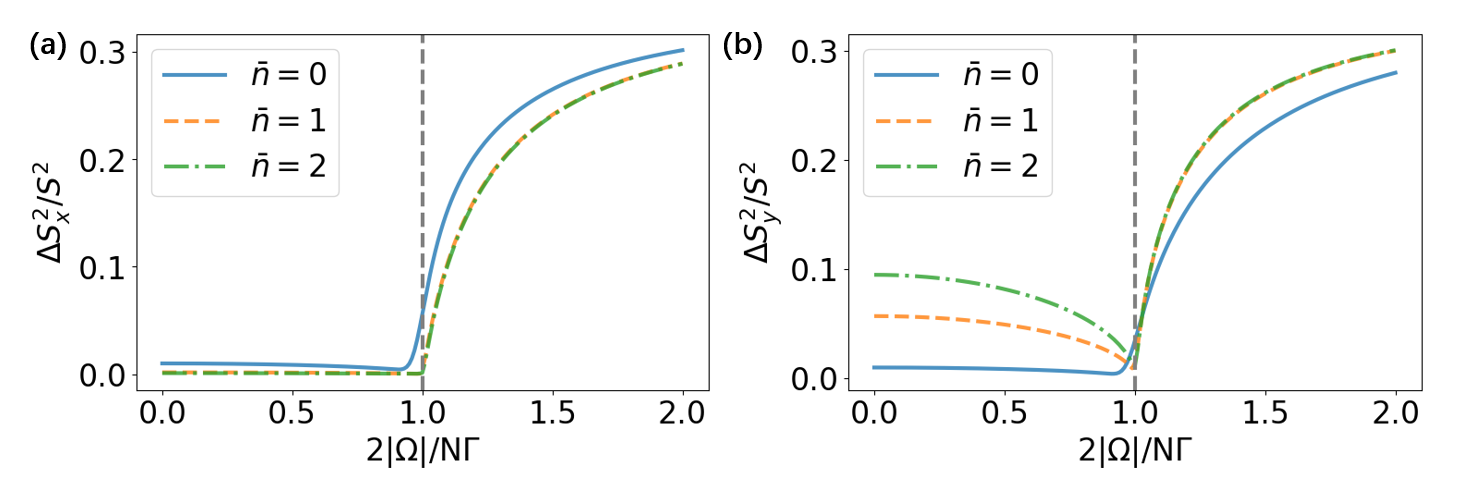} 
    \caption{Fluctuations of the steady state as a function of the scaled drive strength $2|\Omega|/N \Gamma$ for different values of the squeezing parameter $\bar{n}$. (a) Variance of the $x$-component of the collective spin, $\Delta S_x^2$; (b) variance of the $y$-component, $\Delta S_y^2$. The atom number is fixed at $N=100$. }
    \label{phase2}
\end{figure}

As the squeezing strength increases, a characteristic arc-shaped region of negativity appears in the Wigner distribution in the stationary phase. This feature reflects the emergence of atomic squeezing and nonclassical correlations. In both phases, the distribution clearly exhibits enhanced spreading along the $y$-direction, indicating anisotropic quantum noise. These features are quantitatively supported by the fluctuation analysis in Fig. \ref{phase2}, which shows how the fluctuations of transverse spin components depend on the squeezing strength.

Figure \ref{phase2} shows the variances $\Delta S_x^2$ and $\Delta S_y^2$ as functions of the scaled drive strength $2|\Omega|/N\Gamma$. In both phases, squeezing suppresses fluctuations along the $x$-direction, while enhancing them along $y$, consistent with the anisotropic noise introduced by the squeezed reservoir. Fluctuations in both directions grow substantially in the time-crystalline phase relative to the stationary phase. Moreover, squeezing continues to sharpen the dissipative phase transition, as evidenced by the pronounced changes in transverse spin fluctuations near the critical point.
\begin{figure}[h]
    \includegraphics[width=0.45\textwidth]{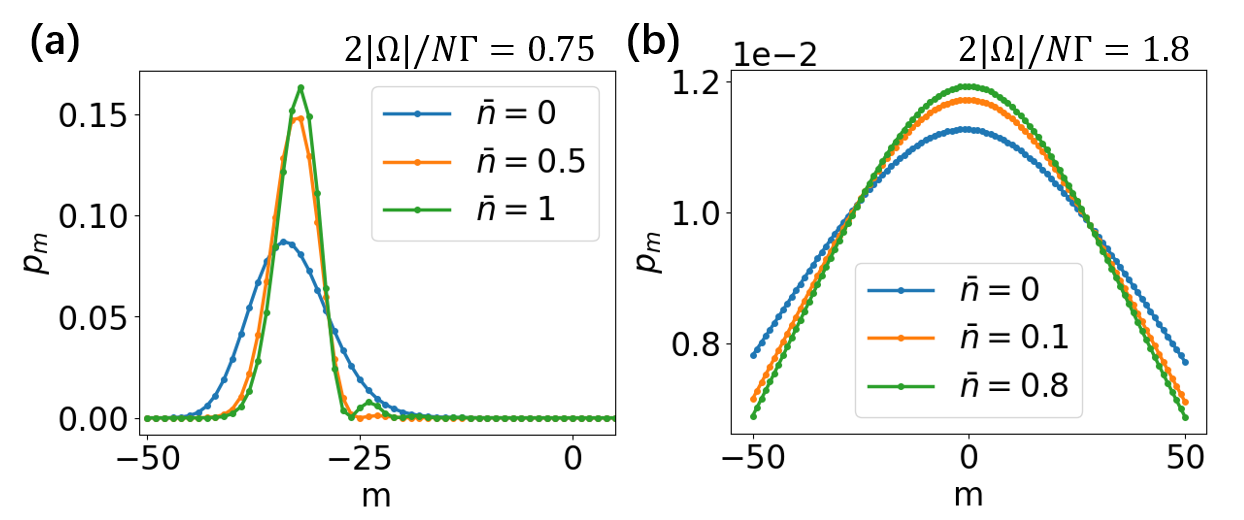}
    \caption{Steady-state occupation probabilities $p_m$ of the collective spin states $\left | m \right >$ for $N=100$ and various values of the squeezing parameter $\bar{n}$. (a) Stationary phase at $2|\Omega|/N \Gamma=0.75$. (b) Time-crystalline phase at $2|\Omega|/N \Gamma=1.8$.}
    \label{pm}
\end{figure}

To further investigate the role of squeezing in the dissipative phase transition, we analyze the steady-state occupation probabilities $p_m$ of the system’s eigenstates. Figure \ref{pm} shows these distributions across different phases and squeezing strengths. In the stationary phase, the distribution is sharply peaked, with the majority of states unoccupied, indicating minimal quantum fluctuations and a highly localized steady state. As the squeezing strength increases, the main peak becomes more pronounced, and a secondary peak emerges, gradually gaining weight. In contrast, the time-crystalline phase exhibits a significantly broader distribution, with substantial occupation across a wide range of states—reflecting enhanced fluctuations and the presence of long-lived coherent dynamics. Interestingly, increased squeezing in this regime leads to a narrowing of the distribution.  

\section{Conclusion}
In conclusion, we have investigated the influence of a squeezed reservoir on dissipative phase transitions and time-crystalline behavior in a driven collective-spin system. Starting from the established framework of open quantum dynamics, we derived a Lindblad master equation describing collective coupling to a broadband squeezed vacuum. Analysis of the steady-state polarization as a function of the scaled drive strength $2|\Omega|/N\Gamma$ reveals that increasing the squeezing strength sharpens the dissipative phase transition. 

Spectral analysis of the Liouvillian superoperator demonstrates the closing of both the first and second dissipative gaps, which serve as clear indicators of the transition. We further identified a nonmonotonic dependence of the Liouvillian gap on the squeezing parameter. This U-shaped behavior highlights the existence of an optimal squeezing regime that supports longer-lived oscillations. The emergence of eigenvalues with vanishing real parts and finite imaginary components confirms the emergence of non-decaying oscillatory modes, signaling persistent time-crystalline order in the thermodynamic limit. As the squeezing parameter $\bar n$ increases, a greater number of such modes emerge among the low-lying Liouvillian eigenstates. Time-domain simulations validate the spectral analysis by demonstrating precise agreement between the observed oscillation frequencies and the imaginary parts of the corresponding Liouvillian eigenvalues.

Finally, we examined the role of squeezing in shaping quantum fluctuations. Wigner function distributions, directional variances, and eigenstate occupation statistics reveal enhanced anisotropy and nonclassical correlations induced by the squeezed reservoir. Together, these results establish a pathway for controlling and diagnosing time-crystalline phases in open quantum systems through engineered dissipation.

\section{Acknowledgments}
We are grateful for the support of Air Force Office of Scientific Research (Award No FA-9550-20-1-0366) and the Robert A Welch Foundation (Grants No. A-1943-20210327 and No. A-1943-20240404). Z. Z. is supported by the Heep Graduate Fellowship. We acknowledge the use
of QuTiP PYTHON library \cite{johansson2012qutip,JOHANSSON2013QuTiP2}.

\bibliography{refs}
\end{document}